\documentclass[aps,pra,twocolumn,groupedaddress]{revtex4-1}

\usepackage{graphicx}
\usepackage{bm}
\newcommand\fig[1]{Fig.~\ref{fig:#1}}
\newcommand\Fig[1]{Figure~\ref{fig:#1}}

\begin{document}


\title{Sisyphus Cooling of Lithium}


\author{Paul Hamilton}
\email[]{paul.hamilton@berkeley.edu}
\author{Geena Kim}
\author{Trinity Joshi}
\author{Biswaroop Mukherjee}
\author{Daniel Tiarks}
\altaffiliation{Present address: Max-Planck Institut f\"{u}r Quantenoptik, Garching, Germany}
\author{Holger M\"{u}ller}
\altaffiliation{Lawrence Berkeley National Laboratory, One Cyclotron Road, Berkeley, California 94720, USA}
\affiliation{Department of Physics, University of California, Berkeley, California 94720, USA}

\date{\today}

\begin{abstract}
Laser cooling to sub-Doppler temperatures by optical molasses is thought to be inhibited in atoms with unresolved, near-degenerate hyperfine structure in the excited state.  We demonstrate that such cooling is possible in one to three dimensions, not only near the standard \emph{D}$_2$ line for laser cooling, but over a wide range extending to the \emph{D}$_1$ line.  Via a combination of Sisyphus cooling followed by adiabatic expansion, we reach temperatures as low as 40$\,\mu$K, which corresponds to atomic velocities a factor of 2.6 above the limit imposed by a single photon recoil.  Our method requires modest laser power at a frequency within reach of standard frequency locking methods.  It is largely insensitive to laser power, polarization and detuning, magnetic fields, and initial hyperfine populations.  Our results suggest that optical molasses should be possible with all alkali species.  
\end{abstract}


\maketitle



Sisyphus cooling of neutral atoms is vital for reaching the ultracold temperatures needed in experiments ranging from metrology \cite{Parker2009} to quantum information \cite{Deutsch2000}.  It is simple to apply for species with resolved hyperfine structure, e.g. sodium \cite{Lett1988}, cesium \cite{Salomon1990}, and rubidium \cite{Shang1991}, which have thus become workhorses in atomic physics.  A new generation of experiments, however, requires atoms offering lighter mass or bosonic and fermionic isotopes.  Achieving sub-Doppler temperatures with these atoms has relied upon sympathetic and/or evaporative cooling - methods that are intrinsically lossy, require timescales of seconds, and favorable collisional properties - or optical lattices that require high laser intensities \cite{Chen1992} or detunings of several hundred gigahertz \cite{Anderson1996}.  Sisyphus cooling has been demonstrated with potassium \cite{Gokhroo2011,Landini2011}, which has partially resolved hyperfine structure, but the same method does not apply to lithium, which has inverted and unresolved hyperfine structure \cite{Landini2011}.  Standard Sisyphus cooling of lithium, which has not been demonstrated \cite{Mewes1999,Denschlag1999,Taglieber2006,Tiecke2009,Wang2007,Duarte2011},  would open the door to simpler, faster, and more efficient experiments in ultracold chemistry, quantum gas microscopy, quantum simulation, and tests of the equivalence principle \cite{Deh2008,Heo2012,Khramov2012,Kraft2006,Marzok2009,Voigt2009,Sherson2010,Bakr2009,Hamilton2012}.  Recently, a sub-Doppler cooling scheme for lithium has been demonstrated using a bichromatic lattice and enhancement from a $\Lambda$-type level structure \cite{Grier2013}.  Here, we demonstrate simple, efficient (up to $\sim\!45\%$ cooled fraction), Sisyphus cooling of lithium to temperatures as low as 40$\,\mu$K in one dimension to 100$\,\mu$K in three dimensions using polarization-gradient cooling beams with a detuning of 1-9\,GHz.  These detunings can be produced using a standard offset laser lock. The cooling process operates on a timescale of milliseconds and requires only modest laser power.  It can thus be integrated easily into experiments using existing diode lasers.   

\begin{figure*}
\includegraphics[scale=0.45]{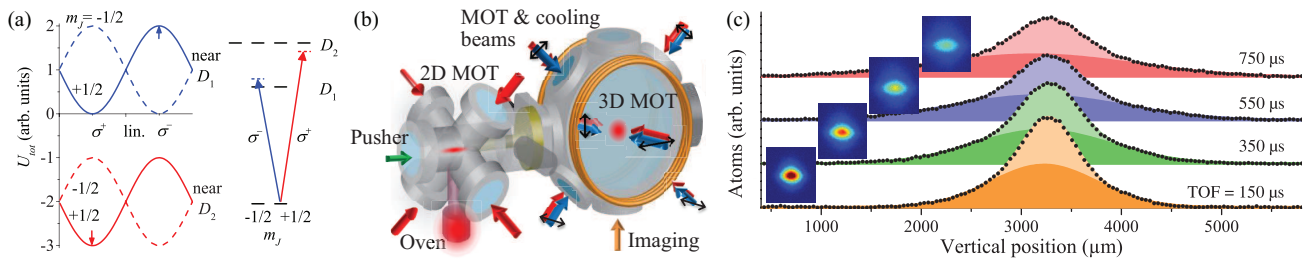}%
\caption{\label{fig:exptsequence}(Color online) Experimental methods.  (a)  Optical potentials and energy levels for Sisyphus cooling.  Left - Solid (dashed) lines indicate the optical potential for the ground state $m_J = +1/2 (-1/2)$ sublevel as a function of the spatially dependent polarization in a lin$\perp$lin standing wave.  Arrows indicate the position of largest Stark shift, indicated for two cases on the right, with detunings near the \emph{D}$_1$ and \emph{D}$_2$ transitions, respectively.  Right - The cooling lattice is detuned by at least 1\,GHz from resonance between the \emph{D}$_1$ and \emph{D}$_2$ transitions which are separated by 10 GHz.    (b)   Setup.  Cooling lattice beams are overlapped with the 3D MOT producing polarization gradients for Sisyphus cooling.  Each counterpropagating pair has orthogonal linear polarizations (black arrows).  (c) Temperature determination using time-of-flight method.  Insets:  Absorption images showing one-dimensional cooling along the vertical axis.  Plots:  Black points are summed data values along the vertical axis.  The light shaded areas are fits to the data using a sum of two Gaussians.  The dark shaded areas indicate the wider Gaussian in each fit.}
\end{figure*}

Historically, when lithium was first laser cooled, sub-Doppler temperatures could not be reached via the standard method of optical molasses slightly red-detuned from the Doppler cooling transition \cite{Schunemann1998,Lin1991}.  The unresolved hyperfine structure of the excited state was thought to prevent the optical pumping needed for Sisyphus cooling \cite{Lin1991,Fort1998,Xu2003}.  However, at detunings large compared to the hyperfine structure [see \fig{exptsequence}(a)], the atomic response is dominated instead by the fine structure.  In fact, in the textbook explanation \cite{Dalibard1989} for Sisyphus cooling, only the fine structure is considered.  Usually these models focus on the 2S$_{1/2}\!\rightarrow\,$2P$_{3/2}$ transition (\emph{D}$_2$) used for initial laser cooling in a magneto-optical trap.  Less well-known is that the Sisyphus mechanism also works with a blue detuning to the 2S$_{1/2}\!\rightarrow\,$2P$_{1/2}$ transition (\emph{D}$_1$) \cite{Guo1993}.  Because the fine structure splitting of the 2P state of lithium is small (10.05\,GHz, as compared to 16.6\,THz for the 6P state of cesium), both transitions contribute constructively and we are able to demonstrate Sisyphus cooling for detunings ranging from \emph{D}$_1$ to \emph{D}$_2$.

\Fig{exptsequence}(b) gives an overview of the apparatus.  Atoms from a hot atomic beam generated by a lithium oven at 400$^\circ$C are captured in a two-dimensional magneto-optical trap (2D MOT) based on the design in \cite{Tiecke2009}, consisting of two retro-reflected cooling beams in the $\sigma^+ / \sigma^-$ configuration and a pusher beam. The 2D MOT beams have a $1/e^2$ intensity radius (``waist") of 5\,mm and a peak intensity of 150\,mW/cm$^2$, while the pusher beam has a waist of 1.2\,mm and a peak intensity of 10\,mW/cm$^2$.  A magnetic field gradient of 50\,G/cm is generated by permanent magnets. The transversely cooled atomic beam passes through a differential pumping stage and is loaded into a 3D MOT with six 6.5\,mm waist beams having a 10\,mW/cm$^2$ peak intensity in the $\sigma^+ / \sigma^-$ configuration.  Approximately equal intensities of the pump and repump frequencies are used in both the 2D MOT and 3D MOT beams.  The sub-Doppler cooling beams are derived from an external cavity diode laser phase locked to the master laser with a frequency offset that can be varied from $-$9 to $+$9\,GHz.  After amplification by a tapered amplifier up to 180\,mW is available for the cooling beams which have 0.7\,mm waists.

The sub-Doppler cooling procedure begins after a one-second loading of 2$\times$10$^7$ $^7$Li atoms at $\sim\!1\,$mK in the 3D MOT.  This is followed by a 5\,ms compressed MOT (CMOT) phase which decreases the temperature to 300-400\,$\mu$K by reducing the pump (repump) power to 2\,mW/cm$^2$ (1.5\,mW/cm$^2$), moving the detuning closer to resonance to $-12\,$MHz ($-11\,$MHz), and increasing the magnetic field gradient from $15\,$G/cm to $30\,$G/cm.  Sisyphus cooling is implemented with a polarization gradient lattice turned on during the last 1\,ms of the CMOT.  In the 3D case, we use three linearly polarized retroreflected beams:  one beam is directed along the axis of the MOT coils, while the other two are overlapped with the MOT beams in the plane of the MOT coils.  The CMOT light is then extinguished and the magnetic field gradient turned off.  The magnetic field decays with a $1/e$ time constant of $\sim\!500\,\mu$s due to eddy currents in the vacuum chamber.  The cooling lattice beams are left on for 500-1500\,$\mu$s after the MOT turnoff.

Alignment of the lattice with both the MOT and magnetic zero is crucial to achieve cooling.  To find magnetic zero, we modulate the MOT magnetic field gradient and iteratively align the MOT cooling beams until there is no discernible movement of the atomic cloud.  The lattice is then aligned, one axis at a time, by first using near-resonant light at low power and observing the disturbance of the MOT.  After this rough alignment, we optimize the final temperature along each axis independently.  Finally, we iteratively adjust the overlap of the three axes to optimize the three-dimensional (3D) cooling.  We have also used ``imaging'' of the cooling beams for alignment \cite{Kerman2002} by first optically pumping the atoms into the $F=1$ hyperfine state and observing the population transfer of the atoms to the $F=2$ by the lattice beams with near resonant light.  We find that the lowest temperatures are achieved when the retroreflected beams are misaligned by a few degrees, which reduces fluctuations caused by reflections off the vacuum windows.

\begin{figure}
\includegraphics[scale=0.75]{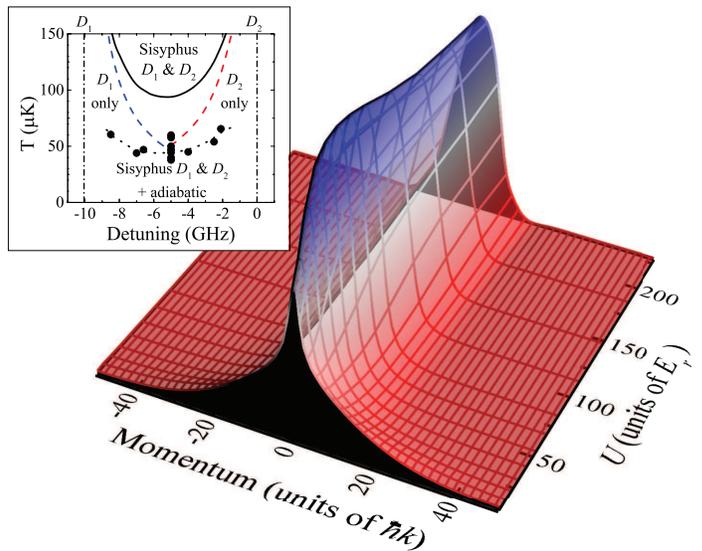}
\caption{\label{fig:oberesults}(Color online) Equilibrium momentum distribution versus optical potential depth, $U$, at a detuning of $\delta =$ -2\,GHz.  At small $U$ the momentum distribution clearly has a narrow peak on a broad background.  For large $U$ the distribution becomes well described by a Gaussian.  Inset:  Comparison of 1D cooling data with theory.  All points are at a resonant Rabi frequency, $\Omega = 32\Gamma$.  At $\delta =$ -2\,GHz this corresponds to $U = 91 E_R$.  The solid line shows the theoretical prediction for Sisyphus cooling only, red(blue) dashed lines show the same predictions with only the \emph{D}$_2$(\emph{D}$_1$) line, and the dotted line includes adiabatic turnoff of the cooling lattice.}
\end{figure}

For measuring the temperature of the cooled atoms, and to determine the fraction of atomic sample cooled, we use the time-of-flight (TOF) method \cite{Ketterle1999}.  The atomic sample is imaged on a CCD camera.  For one- and two-dimensional (1D and 2D) cooling the images are taken perpendicular to the cooling lasers.  To confirm 3D cooling the images were taken at 45 degrees to the in-plane cooling beams and 90 degrees to the on-axis cooling beam [see \fig{exptsequence}(b)].  Cooling is observed along both image axes only when all three dimensions of the atomic sample are cooled.  For TOFs $<$2\,ms we determine the atomic density profile by measuring the absorption of a short ($25-50\,\mu$s) pulse of light resonant with the \emph{D}$_2$ line from the \emph{F}=2 manifold of the ground state.  In order to increase signal to noise for TOF sequences longer than $2\,$ms we use the MOT beams tuned on resonance (100-500\,$\mu$s pulse length) and collect fluorescence.  The observed density profiles have clearly visible cooled and uncooled populations which are fit to a sum of two Gaussians [see \fig{exptsequence}(c)].  Both theoretical calculations \cite{Castin1990} (see \fig{oberesults}) and previous experiments \cite{Jersblad2004} have shown that Sisyphus cooling leads to such a bimodal distribution, particularly at smaller optical potential depths.  The temperature is determined via $k_B T = M \sigma_v^2 = M [\sigma_x^2(t)-\sigma_x^2(0)]/t^2$, where \emph{M} is the atomic mass, $\sigma_v$ is $1/e$ of the velocity distribution, and $\sigma_x(t)$ is the $1/e$ cloud width at time \emph{t}.  The cooled fraction is given by the ratio of the area of the narrow Gaussian to the total area.


\begin{table}
\caption{\label{tab:results}Summary of optimal cooling results.  Intensities for 1D (3D) cooling beams are $5.1(4.3)\times10^3$\,mW/cm$^2$.  See \fig{theoryndata} for the detuning regions.}
\begin{ruledtabular}
\begin{tabular}{ccccc}
Type & Polari- & Region & Temp. ($\mu$K)  & Fraction\\
     & zation         &        & @ Detuning & \\
\hline
3D MOT & $\sigma^+/\sigma^-$ & -35 MHz & 1000 & 100\% \\
3D CMOT & $\sigma^+/\sigma^-$ & -12 MHz & 300 & 100\% \\
\hline
1D Sisyphus & lin$\perp$lin & (b) & 40 {@} -5\,GHz & 25-35\% \\
\& adia.    &               &     &     &         \\
3D Sisyphus & lin$\perp$lin & (b) & 100 {@} -3\,GHz & 30-45\% \\
\& adia.    &               &     &     &         \\
1D adiabatic & lin$\parallel$lin & (a) \& (c) & 10 {@} -13\,GHz & 5-15\% \\
3D adiabatic & lin$\parallel$lin & (a) \& (c) & 60 {@} +2\,GHz & 10-20\%\\
\end{tabular}
\end{ruledtabular}

\end{table}

We achieve temperatures as low as 40\,$\mu$K ($\bar{v} \sim2.6\,v_r$, where $v_r = \hbar k/M$ is the recoil velocity) in 1D cooling and 100\,$\mu$K ($\bar{v} \sim4\,v_r$) in 3D cooling.  The fraction of atoms cooled from the MOT varies from 25-45\% (see Table~\ref{tab:results}).  We find that the cooling is robust with regards to most experimental parameters.  In 1D cooling we achieve the lowest temperature (40\,$\mu$K) and fraction cooled (30-40$\%$) with lin$\perp$lin polarization.  The angle between the polarizations is not critical.  While lin$\parallel$lin shows no cooling, misalignments of as much as 10 degrees from lin$\perp$lin produce show no measurable differences.  In fact, in 3 dimensions, we can even achieve cooling with one axis having lin$\parallel$lin polarization, albeit with slightly higher (150\,$\mu$K) temperature.  In 3D cooling, we also find that the relative polarization between pairs of counterpropagating beams has no effect on the final temperature.  Lowest temperatures are reached with a minimum of 1\,ms temporal overlap of the cooling lattice and the end of the CMOT phase and 1\,ms of cooling lattice after the CMOT is turned off.  This indicates that the cooling process occurs in $<$2\,ms, a timescale consistent with our theoretical calculations.  Leaving the cooling lattice on during loading of the MOT has no effect except when the detuning is small enough ($<$1\,GHz) to disturb the MOT.  The cooling process is insensitive to the initial hyperfine state as optically pumping into either the \emph{F}=1 or \emph{F}=2 hyperfine manifolds during the CMOT phase has no effect on the final temperature or captured fraction.  Using state-selective absorption imaging, we detect $<10\%$ change in the relative populations after the cooling process.  The magnitude of the CMOT magnetic field gradient has a slight effect on the final temperature.  In 1D cooling the final temperature varied from 120\,$\mu$K to 40\,$\mu$K for gradients from 30 to 10\,G/cm with lower temperatures at the lowest gradient.  Artificially increasing the decay time constant of the magnetic field gradient does not change the final temperature but reduces the cooled fraction, as atoms leave the cooling beams.  We also find that constant background magnetic fields ($\sim\!$100\,mG) have little effect on the cooling process as long as the MOT and cooling lattice are well overlapped spatially.  This is likely because the lattice depths used here are much larger than the Zeeman shift in contrast to the optical molasses typically used with other alkali atoms.


We achieve cooling with detunings covering nearly the entire range from the \emph{D}$_1$ to \emph{D}$_2$ transitions, with final temperatures largely independent of the both detuning and laser power.  To compare with theory we investigate 1D cooling over a range of lattice depths and find that our final temperatures of 40-60\,$\mu$K are much lower than the values predicted by the semi-classical theory of Sisyphus cooling \cite{Dalibard1989}.  In addition semi-classical theory predicts that, in the limit of large detuning, the final temperature is universally proportional to the optical potential $U$,
\begin{equation} 
T\propto U=\frac{\hbar\Omega^2}{3\delta},
\label{eq:semiclassical}
\end{equation}
over a range of Rabi frequencies $\Omega$, determined by the laser intensity and detuning $\delta$.  This universality is observed in our data and gave a first clue to the cooling mechanism.

Two additional factors help explain the lower temperatures we observe experimentally.  First, the semi-classical theory calculates an average force on the atom's center-of-mass.  This approximation breaks down once the atomic velocities near the recoil velocity, when the atomic wavepackets become localized in the wells of the optical lattice.  The solid line in the inset of \fig{oberesults} shows the results of a full quantum calculation discretizing the momentum in units of the recoil momentum, $\hbar k$, that we calculated following the model presented in Ref.\,\cite{Castin1990}.  For calculations near the \emph{D}$_2$ (\emph{D}$_1$) line up to 50 external momentum states were used with each of the six (four) internal atomic states, two for the \textit{J}=1/2 ground state and four (two) for the \textit{J}=3/2 (\textit{J}=1/2) excited state.  The coefficients for the optical Bloch equations of the 90000 (40000) possible coherences were then calculated and stored in a sparse matrix.  Equilibrium momentum distributions were determined by either numerically integrating the optical Bloch equations until a steady state was reached or by setting the time derivative of the density matrix to zero and solving the resulting matrix equation.  One finds that at low Rabi frequencies the final momentum distribution consists of a narrow peak with a large tail (see \fig{oberesults}).  The effective temperature is determined by fitting this momentum distribution to two Gaussians, as is done in the experiment, and calculating the equivalent temperature for the narrow peak. The narrow peak consists of $>50\%$ of the atoms for all but the lowest optical potential values ($U<50\,E_r$).  

Second, we find that adiabatic expansion during the turnoff of the polarization gradient lattice reduces the final temperature at higher potential depths \cite{Kastberg1995,Winoto1999}.  The light mass of lithium is advantageous here as it leads to high trap frequencies.  Adiabaticity is reached even with a turnoff time as short as a microsecond. 

The dashed line in the inset of \fig{oberesults} shows the results of the quantum calculation plus the effects of adiabatic expansion using the simple model in \cite{Kastberg1995}.  In this model of adiabatic expansion the final momentum depends only on the vibrational band the atom occupies in the optical lattice.  At a given initial temperature $T_i$, a larger optical potential gives a higher trap frequency, $\omega = 2 E_r\sqrt{U/E_r}/\hbar$, and a larger fraction of the atoms in the ground state as determined by the Boltzmann factor $f_B=e^{-\hbar\omega/k_B T}$.  The final temperature $T_f$ after adiabatic expansion is given by Eq. 2 of~\cite{Kastberg1995},
\begin{equation} 
T_f = T_r\left(\frac{Q_0}{k}\right)^2 \frac{1+4f_B+f_B^2}{12(1-f_B)^2},
\label{eq:adiabatic}
\end{equation}
where $Q_0$ is the lattice constant and $T_r = \frac{\hbar^2 k^2}{k_B M}$ is the recoil temperature.  Adiabatic expansion lowers the final temperature and flattens its change with respect to the optical potential depth in qualitative agreement with our experimental observations.  We have observed that rapid ($<$150\,ns) switching off of the lattice increases the final temperature.  At -5\,GHz detuning, e.g., the temperature increases from 40\,$\mu$K to 90\,$\mu$K, which agrees well with this model.

\begin{figure}
\includegraphics[scale=0.38]{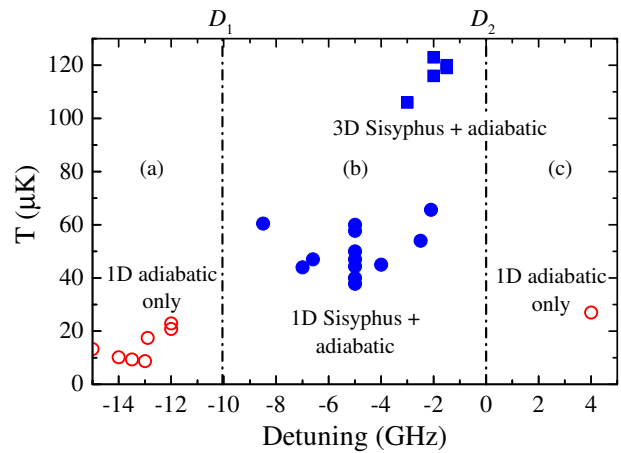}
\caption{\label{fig:theoryndata}(Color online) Final temperature versus the detuning of the cooling lattice.  Circles are 1D cooling, squares are 3D cooling.  Intensities for 1D (3D) cooling beams are $5.1(4.3)\times10^3$\,mW/cm$^2$.  Points at the optimum detuning, -5\,GHz, were taken over several months and demonstrate the robustness of the cooling process.  Solid blue points, between \emph{D}$_1$ and \emph{D}$_2$, use lin$\perp$lin polarization which gives both Sisyphus cooling and adiabatic expansion.  Open red points, outside \emph{D}$_1$ and \emph{D}$_2$, use lin$\parallel$lin polarization which gives only adiabatic cooling.  Lower temperatures are obtained at the expense of a much smaller cooled fraction ($5-15\%$).}
\end{figure}


In summary, we have demonstrated Sisyphus cooling of lithium between the \emph{D}$_1$ and \emph{D}$_2$ transitions followed by adiabatic expansion of the cooling lattice.  The cooling is largely insensitive to most experimental parameters including the precise polarization of the lattice beams, laser power, magnetic fields, initial hyperfine state, timing of the cooling sequence, and detuning of the laser frequency. 

There are improvements that could simplify the setup even further.  Conversion of our six-beam lattice to a four-beam lattice could reduce sensitivity to alignment and vibrations \cite{Grynberg1993}, although our current setup is not overly sensitive.  One could eliminate the additional laser required for the cooling lattice by operating the MOT with lin$\perp$lin polarization (which we have successfully demonstrated) and sweeping the MOT laser from on resonance to the gigahertz detuning needed for sub-Doppler cooling.  Further cooling could be achieved by a stage of Raman sideband cooling in the lattice, which requires one additional optical pumping laser on the \emph{D}$_1$ transition and application of a constant magnetic field \cite{Kerman2000}.

We expect that this technique will find wide applicability in other experiments. For example, the simple implementation and low loss could be useful for experiments using optical lattices where normal molasses cooling is unavailable.  For those experiments requiring temperatures near the recoil limit, our method eliminates the need for either the magnetic trap or high powered lasers required for sympathetic and evaporative cooling.  For experiments which must reach lower temperatures (e.g. Bose or Fermi degenerate gases), our method provides a simple intermediate cooling step which can enhance the loading rate and reduce the depth requirement of traps used for the final stage of cooling.  Thermalization between cooled and uncooled atoms can be avoided by using an initially shallow trap which captures only the cooled population.  The timescale for thermalization is much longer than the time it takes for the hot untrapped atoms to leave such a trap.  Finally, for our purposes, a final temperature of 40\,$\mu$K, which corresponds to an r.m.s velocity of $2.6\,v_r$, will enable the use of a large fraction of our atomic cloud in a first demonstration of interferometry with ultracold lithium.  

\begin{acknowledgments}
We thank Dennis Schlippert and Marcus Ossiander for their contributions to the set up of the apparatus; Michael Hohensee for guidance on the simulations; the groups of Markus Greiner and Dan Stamper-Kurn for useful discussions; and the generous support of the David and Lucile Packard Foundation, the National Aeronautics and Space Administration, and the National Science Foundation. 
\end{acknowledgments}

\bibliography{LiSisyphus}

\begin{thebibliography}{39}%
\makeatletter
\providecommand \@ifxundefined [1]{%
 \@ifx{#1\undefined}
}%
\providecommand \@ifnum [1]{%
 \ifnum #1\expandafter \@firstoftwo
 \else \expandafter \@secondoftwo
 \fi
}%
\providecommand \@ifx [1]{%
 \ifx #1\expandafter \@firstoftwo
 \else \expandafter \@secondoftwo
 \fi
}%
\providecommand \natexlab [1]{#1}%
\providecommand \enquote  [1]{``#1''}%
\providecommand \bibnamefont  [1]{#1}%
\providecommand \bibfnamefont [1]{#1}%
\providecommand \citenamefont [1]{#1}%
\providecommand \href@noop [0]{\@secondoftwo}%
\providecommand \href [0]{\begingroup \@sanitize@url \@href}%
\providecommand \@href[1]{\@@startlink{#1}\@@href}%
\providecommand \@@href[1]{\endgroup#1\@@endlink}%
\providecommand \@sanitize@url [0]{\catcode `\\12\catcode `\$12\catcode
  `\&12\catcode `\#12\catcode `\^12\catcode `\_12\catcode `\%12\relax}%
\providecommand \@@startlink[1]{}%
\providecommand \@@endlink[0]{}%
\providecommand \url  [0]{\begingroup\@sanitize@url \@url }%
\providecommand \@url [1]{\endgroup\@href {#1}{\urlprefix }}%
\providecommand \urlprefix  [0]{URL }%
\providecommand \Eprint [0]{\href }%
\providecommand \doibase [0]{http://dx.doi.org/}%
\providecommand \selectlanguage [0]{\@gobble}%
\providecommand \bibinfo  [0]{\@secondoftwo}%
\providecommand \bibfield  [0]{\@secondoftwo}%
\providecommand \translation [1]{[#1]}%
\providecommand \BibitemOpen [0]{}%
\providecommand \bibitemStop [0]{}%
\providecommand \bibitemNoStop [0]{.\EOS\space}%
\providecommand \EOS [0]{\spacefactor3000\relax}%
\providecommand \BibitemShut  [1]{\csname bibitem#1\endcsname}%
\let\auto@bib@innerbib\@empty
\bibitem [{\citenamefont {Parker}(2009)}]{Parker2009}%
  \BibitemOpen
  \bibfield  {author} {\bibinfo {author} {\bibfnamefont {T.}~\bibnamefont
  {Parker}},\ }\href {http://dx.doi.org/10.1088/0026-1394/47/1/001} {\bibfield
  {journal} {\bibinfo  {journal} {Metrologia}\ }\textbf {\bibinfo {volume}
  {47}},\ \bibinfo {pages} {1} (\bibinfo {year} {2009})}\BibitemShut {NoStop}%
\bibitem [{\citenamefont {Deutsch}\ \emph {et~al.}(2000)\citenamefont
  {Deutsch}, \citenamefont {Brennen},\ and\ \citenamefont
  {Jessen}}]{Deutsch2000}%
  \BibitemOpen
  \bibfield  {author} {\bibinfo {author} {\bibfnamefont {I.}~\bibnamefont
  {Deutsch}}, \bibinfo {author} {\bibfnamefont {G.}~\bibnamefont {Brennen}}, \
  and\ \bibinfo {author} {\bibfnamefont {P.}~\bibnamefont {Jessen}},\ }\href
  {http://dx.doi.org/10.1002/1521-3978(200009)48:9/11%3C925::AID-PROP925%3E3.0.CO;2-A}
  {\bibfield  {journal} {\bibinfo  {journal} {Fortschr. Phys.}\ }\textbf
  {\bibinfo {volume} {48}},\ \bibinfo {pages} {925} (\bibinfo {year}
  {2000})}\BibitemShut {NoStop}%
\bibitem [{\citenamefont {Lett}\ \emph {et~al.}(1988)\citenamefont {Lett},
  \citenamefont {Watts}, \citenamefont {Westbrook}, \citenamefont {Phillips},
  \citenamefont {Gould},\ and\ \citenamefont {Metcalf}}]{Lett1988}%
  \BibitemOpen
  \bibfield  {author} {\bibinfo {author} {\bibfnamefont {P.~D.}\ \bibnamefont
  {Lett}}, \bibinfo {author} {\bibfnamefont {R.~N.}\ \bibnamefont {Watts}},
  \bibinfo {author} {\bibfnamefont {C.~I.}\ \bibnamefont {Westbrook}}, \bibinfo
  {author} {\bibfnamefont {W.~D.}\ \bibnamefont {Phillips}}, \bibinfo {author}
  {\bibfnamefont {P.~L.}\ \bibnamefont {Gould}}, \ and\ \bibinfo {author}
  {\bibfnamefont {H.~J.}\ \bibnamefont {Metcalf}},\ }\href {\doibase
  10.1103/PhysRevLett.61.169} {\bibfield  {journal} {\bibinfo  {journal} {Phys.
  Rev. Lett.}\ }\textbf {\bibinfo {volume} {61}},\ \bibinfo {pages} {169}
  (\bibinfo {year} {1988})}\BibitemShut {NoStop}%
\bibitem [{\citenamefont {Salomon}\ \emph {et~al.}(1990)\citenamefont
  {Salomon}, \citenamefont {Dalibard}, \citenamefont {Phillips}, \citenamefont
  {Clairon},\ and\ \citenamefont {Guellati}}]{Salomon1990}%
  \BibitemOpen
  \bibfield  {author} {\bibinfo {author} {\bibfnamefont {C.}~\bibnamefont
  {Salomon}}, \bibinfo {author} {\bibfnamefont {J.}~\bibnamefont {Dalibard}},
  \bibinfo {author} {\bibfnamefont {W.}~\bibnamefont {Phillips}}, \bibinfo
  {author} {\bibfnamefont {A.}~\bibnamefont {Clairon}}, \ and\ \bibinfo
  {author} {\bibfnamefont {S.}~\bibnamefont {Guellati}},\ }\href
  {http://dx.doi.org/10.1209/0295-5075/12/8/003} {\bibfield  {journal}
  {\bibinfo  {journal} {Europhys. Lett.}\ }\textbf {\bibinfo {volume} {12}},\
  \bibinfo {pages} {683} (\bibinfo {year} {1990})}\BibitemShut {NoStop}%
\bibitem [{\citenamefont {Shang}\ \emph {et~al.}(1991)\citenamefont {Shang},
  \citenamefont {Sheehy}, \citenamefont {Metcalf}, \citenamefont {van~der
  Straten},\ and\ \citenamefont {Nienhuis}}]{Shang1991}%
  \BibitemOpen
  \bibfield  {author} {\bibinfo {author} {\bibfnamefont {S.~Q.}\ \bibnamefont
  {Shang}}, \bibinfo {author} {\bibfnamefont {B.}~\bibnamefont {Sheehy}},
  \bibinfo {author} {\bibfnamefont {H.}~\bibnamefont {Metcalf}}, \bibinfo
  {author} {\bibfnamefont {P.}~\bibnamefont {van~der Straten}}, \ and\ \bibinfo
  {author} {\bibfnamefont {G.}~\bibnamefont {Nienhuis}},\ }\href
  {http://dx.doi.org/10.1103/PhysRevLett.67.1094} {\bibfield  {journal}
  {\bibinfo  {journal} {Phys. Rev. Lett.}\ }\textbf {\bibinfo {volume} {67}},\
  \bibinfo {pages} {1094} (\bibinfo {year} {1991})}\BibitemShut {NoStop}%
\bibitem [{\citenamefont {Chen}\ \emph {et~al.}(1992)\citenamefont {Chen},
  \citenamefont {Story}, \citenamefont {Tollett},\ and\ \citenamefont
  {Hulet}}]{Chen1992}%
  \BibitemOpen
  \bibfield  {author} {\bibinfo {author} {\bibfnamefont {J.}~\bibnamefont
  {Chen}}, \bibinfo {author} {\bibfnamefont {J.~G.}\ \bibnamefont {Story}},
  \bibinfo {author} {\bibfnamefont {J.~J.}\ \bibnamefont {Tollett}}, \ and\
  \bibinfo {author} {\bibfnamefont {R.~G.}\ \bibnamefont {Hulet}},\ }\href
  {http://dx.doi.org/10.1103/PhysRevLett.69.1344} {\bibfield  {journal}
  {\bibinfo  {journal} {Phys. Rev. Lett.}\ }\textbf {\bibinfo {volume} {69}},\
  \bibinfo {pages} {1344} (\bibinfo {year} {1992})}\BibitemShut {NoStop}%
\bibitem [{\citenamefont {Anderson}\ \emph {et~al.}(1996)\citenamefont
  {Anderson}, \citenamefont {Gustavson},\ and\ \citenamefont
  {Kasevich}}]{Anderson1996}%
  \BibitemOpen
  \bibfield  {author} {\bibinfo {author} {\bibfnamefont {B.~P.}\ \bibnamefont
  {Anderson}}, \bibinfo {author} {\bibfnamefont {T.~L.}\ \bibnamefont
  {Gustavson}}, \ and\ \bibinfo {author} {\bibfnamefont {M.~A.}\ \bibnamefont
  {Kasevich}},\ }\href {http://dx.doi.org/10.1103/PhysRevA.53.R3727} {\bibfield
   {journal} {\bibinfo  {journal} {Phys. Rev. A}\ }\textbf {\bibinfo {volume}
  {53}},\ \bibinfo {pages} {R3727} (\bibinfo {year} {1996})}\BibitemShut
  {NoStop}%
\bibitem [{\citenamefont {Gokhroo}\ \emph {et~al.}(2011)\citenamefont
  {Gokhroo}, \citenamefont {Rajalakshmi}, \citenamefont {Easwaran},\ and\
  \citenamefont {Unnikrishnan}}]{Gokhroo2011}%
  \BibitemOpen
  \bibfield  {author} {\bibinfo {author} {\bibfnamefont {V.}~\bibnamefont
  {Gokhroo}}, \bibinfo {author} {\bibfnamefont {G.}~\bibnamefont
  {Rajalakshmi}}, \bibinfo {author} {\bibfnamefont {K.}~\bibnamefont
  {Easwaran}}, \ and\ \bibinfo {author} {\bibfnamefont {C.}~\bibnamefont
  {Unnikrishnan}},\ }\href {http://dx.doi.org/10.1088/0953-4075/44/11/115307}
  {\bibfield  {journal} {\bibinfo  {journal} {J. Phys. B}\ }\textbf {\bibinfo
  {volume} {44}},\ \bibinfo {pages} {115307} (\bibinfo {year}
  {2011})}\BibitemShut {NoStop}%
\bibitem [{\citenamefont {Landini}\ \emph {et~al.}(2011)\citenamefont
  {Landini}, \citenamefont {Roy}, \citenamefont {Carcagn\'{i}}, \citenamefont
  {Trypogeorgos}, \citenamefont {Fattori}, \citenamefont {Inguscio},\ and\
  \citenamefont {Modugno}}]{Landini2011}%
  \BibitemOpen
  \bibfield  {author} {\bibinfo {author} {\bibfnamefont {M.}~\bibnamefont
  {Landini}}, \bibinfo {author} {\bibfnamefont {S.}~\bibnamefont {Roy}},
  \bibinfo {author} {\bibfnamefont {L.}~\bibnamefont {Carcagn\'{i}}}, \bibinfo
  {author} {\bibfnamefont {D.}~\bibnamefont {Trypogeorgos}}, \bibinfo {author}
  {\bibfnamefont {M.}~\bibnamefont {Fattori}}, \bibinfo {author} {\bibfnamefont
  {M.}~\bibnamefont {Inguscio}}, \ and\ \bibinfo {author} {\bibfnamefont
  {G.}~\bibnamefont {Modugno}},\ }\href
  {http://dx.doi.org/10.1103/PhysRevA.84.043432} {\bibfield  {journal}
  {\bibinfo  {journal} {Phys. Rev. A}\ }\textbf {\bibinfo {volume} {84}},\
  \bibinfo {pages} {043432} (\bibinfo {year} {2011})}\BibitemShut {NoStop}%
\bibitem [{\citenamefont {Mewes}\ \emph {et~al.}(1999)\citenamefont {Mewes},
  \citenamefont {Ferrari}, \citenamefont {Schreck}, \citenamefont {Sinatra},\
  and\ \citenamefont {Salomon}}]{Mewes1999}%
  \BibitemOpen
  \bibfield  {author} {\bibinfo {author} {\bibfnamefont {M.~O.}\ \bibnamefont
  {Mewes}}, \bibinfo {author} {\bibfnamefont {G.}~\bibnamefont {Ferrari}},
  \bibinfo {author} {\bibfnamefont {F.}~\bibnamefont {Schreck}}, \bibinfo
  {author} {\bibfnamefont {A.}~\bibnamefont {Sinatra}}, \ and\ \bibinfo
  {author} {\bibfnamefont {C.}~\bibnamefont {Salomon}},\ }\href {\doibase
  10.1103/physreva.61.011403} {\bibfield  {journal} {\bibinfo  {journal}
  {Physical Review A}\ }\textbf {\bibinfo {volume} {61}},\ \bibinfo {pages}
  {011403+} (\bibinfo {year} {1999})}\BibitemShut {NoStop}%
\bibitem [{\citenamefont {Denschlag}\ \emph {et~al.}(1999)\citenamefont
  {Denschlag}, \citenamefont {Cassettari},\ and\ \citenamefont
  {Schmiedmayer}}]{Denschlag1999}%
  \BibitemOpen
  \bibfield  {author} {\bibinfo {author} {\bibfnamefont {J.}~\bibnamefont
  {Denschlag}}, \bibinfo {author} {\bibfnamefont {D.}~\bibnamefont
  {Cassettari}}, \ and\ \bibinfo {author} {\bibfnamefont {J.}~\bibnamefont
  {Schmiedmayer}},\ }\href {\doibase 10.1103/physrevlett.82.2014} {\bibfield
  {journal} {\bibinfo  {journal} {Physical Review Letters}\ }\textbf {\bibinfo
  {volume} {82}},\ \bibinfo {pages} {2014} (\bibinfo {year}
  {1999})}\BibitemShut {NoStop}%
\bibitem [{\citenamefont {Taglieber}\ \emph {et~al.}(2006)\citenamefont
  {Taglieber}, \citenamefont {Voigt}, \citenamefont {Henkel}, \citenamefont
  {Fray}, \citenamefont {H\"{a}nsch},\ and\ \citenamefont
  {Dieckmann}}]{Taglieber2006}%
  \BibitemOpen
  \bibfield  {author} {\bibinfo {author} {\bibfnamefont {M.}~\bibnamefont
  {Taglieber}}, \bibinfo {author} {\bibfnamefont {A.~C.}\ \bibnamefont
  {Voigt}}, \bibinfo {author} {\bibfnamefont {F.}~\bibnamefont {Henkel}},
  \bibinfo {author} {\bibfnamefont {S.}~\bibnamefont {Fray}}, \bibinfo {author}
  {\bibfnamefont {T.~W.}\ \bibnamefont {H\"{a}nsch}}, \ and\ \bibinfo {author}
  {\bibfnamefont {K.}~\bibnamefont {Dieckmann}},\ }\href {\doibase
  10.1103/physreva.73.011402} {\bibfield  {journal} {\bibinfo  {journal}
  {Physical Review A}\ }\textbf {\bibinfo {volume} {73}},\ \bibinfo {pages}
  {011402+} (\bibinfo {year} {2006})}\BibitemShut {NoStop}%
\bibitem [{\citenamefont {Tiecke}\ \emph {et~al.}(2009)\citenamefont {Tiecke},
  \citenamefont {Gensemer}, \citenamefont {Ludewig},\ and\ \citenamefont
  {Walraven}}]{Tiecke2009}%
  \BibitemOpen
  \bibfield  {author} {\bibinfo {author} {\bibfnamefont {T.~G.}\ \bibnamefont
  {Tiecke}}, \bibinfo {author} {\bibfnamefont {S.~D.}\ \bibnamefont
  {Gensemer}}, \bibinfo {author} {\bibfnamefont {A.}~\bibnamefont {Ludewig}}, \
  and\ \bibinfo {author} {\bibfnamefont {J.~T.~M.}\ \bibnamefont {Walraven}},\
  }\href {http://dx.doi.org/10.1103/PhysRevA.80.013409} {\bibfield  {journal}
  {\bibinfo  {journal} {Phys. Rev. A}\ }\textbf {\bibinfo {volume} {80}},\
  \bibinfo {pages} {013409} (\bibinfo {year} {2009})}\BibitemShut {NoStop}%
\bibitem [{\citenamefont {Wang}\ \emph {et~al.}(2007)\citenamefont {Wang},
  \citenamefont {Liu}, \citenamefont {Minardi},\ and\ \citenamefont
  {Kasevich}}]{Wang2007}%
  \BibitemOpen
  \bibfield  {author} {\bibinfo {author} {\bibfnamefont {R.}~\bibnamefont
  {Wang}}, \bibinfo {author} {\bibfnamefont {M.}~\bibnamefont {Liu}}, \bibinfo
  {author} {\bibfnamefont {F.}~\bibnamefont {Minardi}}, \ and\ \bibinfo
  {author} {\bibfnamefont {M.}~\bibnamefont {Kasevich}},\ }\href {\doibase
  10.1103/physreva.75.013610} {\bibfield  {journal} {\bibinfo  {journal}
  {Physical Review A}\ }\textbf {\bibinfo {volume} {75}},\ \bibinfo {pages}
  {013610+} (\bibinfo {year} {2007})}\BibitemShut {NoStop}%
\bibitem [{\citenamefont {Duarte}\ \emph {et~al.}(2011)\citenamefont {Duarte},
  \citenamefont {Hart}, \citenamefont {Hitchcock}, \citenamefont {Corcovilos},
  \citenamefont {Yang}, \citenamefont {Reed},\ and\ \citenamefont
  {Hulet}}]{Duarte2011}%
  \BibitemOpen
  \bibfield  {author} {\bibinfo {author} {\bibfnamefont {P.~M.}\ \bibnamefont
  {Duarte}}, \bibinfo {author} {\bibfnamefont {R.~A.}\ \bibnamefont {Hart}},
  \bibinfo {author} {\bibfnamefont {J.~M.}\ \bibnamefont {Hitchcock}}, \bibinfo
  {author} {\bibfnamefont {T.~A.}\ \bibnamefont {Corcovilos}}, \bibinfo
  {author} {\bibfnamefont {T.~L.}\ \bibnamefont {Yang}}, \bibinfo {author}
  {\bibfnamefont {A.}~\bibnamefont {Reed}}, \ and\ \bibinfo {author}
  {\bibfnamefont {R.~G.}\ \bibnamefont {Hulet}},\ }\href
  {http://dx.doi.org/10.1103/PhysRevA.84.061406} {\bibfield  {journal}
  {\bibinfo  {journal} {Phys. Rev. A}\ }\textbf {\bibinfo {volume} {84}},\
  \bibinfo {pages} {061406} (\bibinfo {year} {2011})}\BibitemShut {NoStop}%
\bibitem [{\citenamefont {Deh}\ \emph {et~al.}(2008)\citenamefont {Deh},
  \citenamefont {Marzok}, \citenamefont {Zimmermann},\ and\ \citenamefont
  {Courteille}}]{Deh2008}%
  \BibitemOpen
  \bibfield  {author} {\bibinfo {author} {\bibfnamefont {B.}~\bibnamefont
  {Deh}}, \bibinfo {author} {\bibfnamefont {C.}~\bibnamefont {Marzok}},
  \bibinfo {author} {\bibfnamefont {C.}~\bibnamefont {Zimmermann}}, \ and\
  \bibinfo {author} {\bibfnamefont {P.~W.}\ \bibnamefont {Courteille}},\ }\href
  {http://dx.doi.org/10.1103/PhysRevA.77.010701} {\bibfield  {journal}
  {\bibinfo  {journal} {Phys. Rev. A}\ }\textbf {\bibinfo {volume} {77}},\
  \bibinfo {pages} {010701} (\bibinfo {year} {2008})}\BibitemShut {NoStop}%
\bibitem [{\citenamefont {Heo}\ \emph {et~al.}(2012)\citenamefont {Heo},
  \citenamefont {Wang}, \citenamefont {Christensen}, \citenamefont {Rvachov},
  \citenamefont {Cotta}, \citenamefont {Choi}, \citenamefont {Lee},\ and\
  \citenamefont {Ketterle}}]{Heo2012}%
  \BibitemOpen
  \bibfield  {author} {\bibinfo {author} {\bibfnamefont {M.~S.}\ \bibnamefont
  {Heo}}, \bibinfo {author} {\bibfnamefont {T.~T.}\ \bibnamefont {Wang}},
  \bibinfo {author} {\bibfnamefont {C.~A.}\ \bibnamefont {Christensen}},
  \bibinfo {author} {\bibfnamefont {T.~M.}\ \bibnamefont {Rvachov}}, \bibinfo
  {author} {\bibfnamefont {D.~A.}\ \bibnamefont {Cotta}}, \bibinfo {author}
  {\bibfnamefont {J.~H.}\ \bibnamefont {Choi}}, \bibinfo {author}
  {\bibfnamefont {Y.~R.}\ \bibnamefont {Lee}}, \ and\ \bibinfo {author}
  {\bibfnamefont {W.}~\bibnamefont {Ketterle}},\ }\href
  {http://dx.doi.org/10.1103/PhysRevA.86.021602} {\bibfield  {journal}
  {\bibinfo  {journal} {Phys. Rev. A}\ }\textbf {\bibinfo {volume} {86}},\
  \bibinfo {pages} {021602} (\bibinfo {year} {2012})}\BibitemShut {NoStop}%
\bibitem [{\citenamefont {Khramov}\ \emph {et~al.}(2012)\citenamefont
  {Khramov}, \citenamefont {Hansen}, \citenamefont {Jamison}, \citenamefont
  {Dowd},\ and\ \citenamefont {Gupta}}]{Khramov2012}%
  \BibitemOpen
  \bibfield  {author} {\bibinfo {author} {\bibfnamefont {A.~Y.}\ \bibnamefont
  {Khramov}}, \bibinfo {author} {\bibfnamefont {A.~H.}\ \bibnamefont {Hansen}},
  \bibinfo {author} {\bibfnamefont {A.~O.}\ \bibnamefont {Jamison}}, \bibinfo
  {author} {\bibfnamefont {W.~H.}\ \bibnamefont {Dowd}}, \ and\ \bibinfo
  {author} {\bibfnamefont {S.}~\bibnamefont {Gupta}},\ }\href
  {http://dx.doi.org/10.1103/PhysRevA.86.032705} {\bibfield  {journal}
  {\bibinfo  {journal} {Phys. Rev. A}\ }\textbf {\bibinfo {volume} {86}},\
  \bibinfo {pages} {032705} (\bibinfo {year} {2012})}\BibitemShut {NoStop}%
\bibitem [{\citenamefont {Kraft}\ \emph {et~al.}(2006)\citenamefont {Kraft},
  \citenamefont {Staanum}, \citenamefont {Lange}, \citenamefont {Vogel},
  \citenamefont {Wester},\ and\ \citenamefont {Weidem\"{u}ller}}]{Kraft2006}%
  \BibitemOpen
  \bibfield  {author} {\bibinfo {author} {\bibfnamefont {S.}~\bibnamefont
  {Kraft}}, \bibinfo {author} {\bibfnamefont {P.}~\bibnamefont {Staanum}},
  \bibinfo {author} {\bibfnamefont {J.}~\bibnamefont {Lange}}, \bibinfo
  {author} {\bibfnamefont {L.}~\bibnamefont {Vogel}}, \bibinfo {author}
  {\bibfnamefont {R.}~\bibnamefont {Wester}}, \ and\ \bibinfo {author}
  {\bibfnamefont {M.}~\bibnamefont {Weidem\"{u}ller}},\ }\href
  {http://dx.doi.org/10.1088/0953-4075/39/19/S13} {\bibfield  {journal}
  {\bibinfo  {journal} {J. Phys. B}\ }\textbf {\bibinfo {volume} {39}},\
  \bibinfo {pages} {S993} (\bibinfo {year} {2006})}\BibitemShut {NoStop}%
\bibitem [{\citenamefont {Marzok}\ \emph {et~al.}(2009)\citenamefont {Marzok},
  \citenamefont {Deh}, \citenamefont {Zimmermann}, \citenamefont {Courteille},
  \citenamefont {Tiemann}, \citenamefont {Vanne},\ and\ \citenamefont
  {Saenz}}]{Marzok2009}%
  \BibitemOpen
  \bibfield  {author} {\bibinfo {author} {\bibfnamefont {C.}~\bibnamefont
  {Marzok}}, \bibinfo {author} {\bibfnamefont {B.}~\bibnamefont {Deh}},
  \bibinfo {author} {\bibfnamefont {C.}~\bibnamefont {Zimmermann}}, \bibinfo
  {author} {\bibfnamefont {P.~W.}\ \bibnamefont {Courteille}}, \bibinfo
  {author} {\bibfnamefont {E.}~\bibnamefont {Tiemann}}, \bibinfo {author}
  {\bibfnamefont {Y.~V.}\ \bibnamefont {Vanne}}, \ and\ \bibinfo {author}
  {\bibfnamefont {A.}~\bibnamefont {Saenz}},\ }\href
  {http://dx.doi.org/10.1103/PhysRevA.79.012717} {\bibfield  {journal}
  {\bibinfo  {journal} {Phys. Rev. A}\ }\textbf {\bibinfo {volume} {79}},\
  \bibinfo {pages} {012717} (\bibinfo {year} {2009})}\BibitemShut {NoStop}%
\bibitem [{\citenamefont {Voigt}\ \emph {et~al.}(2009)\citenamefont {Voigt},
  \citenamefont {Taglieber}, \citenamefont {Costa}, \citenamefont {Aoki},
  \citenamefont {Wieser}, \citenamefont {H\"{a}nsch},\ and\ \citenamefont
  {Dieckmann}}]{Voigt2009}%
  \BibitemOpen
  \bibfield  {author} {\bibinfo {author} {\bibfnamefont {A.~C.}\ \bibnamefont
  {Voigt}}, \bibinfo {author} {\bibfnamefont {M.}~\bibnamefont {Taglieber}},
  \bibinfo {author} {\bibfnamefont {L.}~\bibnamefont {Costa}}, \bibinfo
  {author} {\bibfnamefont {T.}~\bibnamefont {Aoki}}, \bibinfo {author}
  {\bibfnamefont {W.}~\bibnamefont {Wieser}}, \bibinfo {author} {\bibfnamefont
  {T.~W.}\ \bibnamefont {H\"{a}nsch}}, \ and\ \bibinfo {author} {\bibfnamefont
  {K.}~\bibnamefont {Dieckmann}},\ }\href
  {http://dx.doi.org/10.1103/PhysRevLett.102.020405} {\bibfield  {journal}
  {\bibinfo  {journal} {Phys. Rev. Lett.}\ }\textbf {\bibinfo {volume} {102}},\
  \bibinfo {pages} {020405} (\bibinfo {year} {2009})}\BibitemShut {NoStop}%
\bibitem [{\citenamefont {Sherson}\ \emph {et~al.}(2010)\citenamefont
  {Sherson}, \citenamefont {Weitenberg}, \citenamefont {Endres}, \citenamefont
  {Cheneau}, \citenamefont {Bloch},\ and\ \citenamefont {Kuhr}}]{Sherson2010}%
  \BibitemOpen
  \bibfield  {author} {\bibinfo {author} {\bibfnamefont {J.}~\bibnamefont
  {Sherson}}, \bibinfo {author} {\bibfnamefont {C.}~\bibnamefont {Weitenberg}},
  \bibinfo {author} {\bibfnamefont {M.}~\bibnamefont {Endres}}, \bibinfo
  {author} {\bibfnamefont {M.}~\bibnamefont {Cheneau}}, \bibinfo {author}
  {\bibfnamefont {I.}~\bibnamefont {Bloch}}, \ and\ \bibinfo {author}
  {\bibfnamefont {S.}~\bibnamefont {Kuhr}},\ }\href
  {http://dx.doi.org/10.1038/nature09378} {\bibfield  {journal} {\bibinfo
  {journal} {Nature (London)}\ }\textbf {\bibinfo {volume} {467}},\ \bibinfo
  {pages} {68} (\bibinfo {year} {2010})}\BibitemShut {NoStop}%
\bibitem [{\citenamefont {Bakr}\ \emph {et~al.}(2009)\citenamefont {Bakr},
  \citenamefont {Gillen}, \citenamefont {Peng}, \citenamefont {F\"{o}lling},\
  and\ \citenamefont {Greiner}}]{Bakr2009}%
  \BibitemOpen
  \bibfield  {author} {\bibinfo {author} {\bibfnamefont {W.}~\bibnamefont
  {Bakr}}, \bibinfo {author} {\bibfnamefont {J.}~\bibnamefont {Gillen}},
  \bibinfo {author} {\bibfnamefont {A.}~\bibnamefont {Peng}}, \bibinfo {author}
  {\bibfnamefont {S.}~\bibnamefont {F\"{o}lling}}, \ and\ \bibinfo {author}
  {\bibfnamefont {M.}~\bibnamefont {Greiner}},\ }\href
  {http://dx.doi.org/10.1038/nature08482} {\bibfield  {journal} {\bibinfo
  {journal} {Nature (London)}\ }\textbf {\bibinfo {volume} {462}},\ \bibinfo
  {pages} {74} (\bibinfo {year} {2009})}\BibitemShut {NoStop}%
\bibitem [{\citenamefont {Hamilton}\ \emph {et~al.}(2012)\citenamefont
  {Hamilton}, \citenamefont {Barter}, \citenamefont {Kim}, \citenamefont
  {Mukherjee},\ and\ \citenamefont {M\"{u}ller}}]{Hamilton2012}%
  \BibitemOpen
  \bibfield  {author} {\bibinfo {author} {\bibfnamefont {P.}~\bibnamefont
  {Hamilton}}, \bibinfo {author} {\bibfnamefont {T.}~\bibnamefont {Barter}},
  \bibinfo {author} {\bibfnamefont {G.}~\bibnamefont {Kim}}, \bibinfo {author}
  {\bibfnamefont {B.}~\bibnamefont {Mukherjee}}, \ and\ \bibinfo {author}
  {\bibfnamefont {H.}~\bibnamefont {M\"{u}ller}},\ }\href
  {http://meetings.aps.org/link/BAPS.2012.DAMOP.T5.4} {\bibfield  {journal}
  {\bibinfo  {journal} {Bull. Am. Phys. Soc.}\ }\textbf {\bibinfo {volume}
  {57}},\ \bibinfo {pages} {T5.00004} (\bibinfo {year} {2012})}\BibitemShut
  {NoStop}%
\bibitem [{\citenamefont {Grier}\ \emph {et~al.}(2013)\citenamefont {Grier},
  \citenamefont {Ferrier-Barbut}, \citenamefont {Rem}, \citenamefont
  {Delehaye}, \citenamefont {Khaykovich}, \citenamefont {Chevy},\ and\
  \citenamefont {Salomon}}]{Grier2013}%
  \BibitemOpen
  \bibfield  {author} {\bibinfo {author} {\bibfnamefont {A.~T.}\ \bibnamefont
  {Grier}}, \bibinfo {author} {\bibfnamefont {I.}~\bibnamefont
  {Ferrier-Barbut}}, \bibinfo {author} {\bibfnamefont {B.~S.}\ \bibnamefont
  {Rem}}, \bibinfo {author} {\bibfnamefont {M.}~\bibnamefont {Delehaye}},
  \bibinfo {author} {\bibfnamefont {L.}~\bibnamefont {Khaykovich}}, \bibinfo
  {author} {\bibfnamefont {F.}~\bibnamefont {Chevy}}, \ and\ \bibinfo {author}
  {\bibfnamefont {C.}~\bibnamefont {Salomon}},\ }\href
  {http://dx.doi.org/10.1103/physreva.87.063411} {\bibfield  {journal}
  {\bibinfo  {journal} {Physical Review A}\ }\textbf {\bibinfo {volume} {87}},\
  \bibinfo {pages} {063411} (\bibinfo {year} {2013})}\BibitemShut {NoStop}%
\bibitem [{\citenamefont {Sch\"{u}nemann}\ \emph {et~al.}(1998)\citenamefont
  {Sch\"{u}nemann}, \citenamefont {Engler}, \citenamefont {Zielonkowski},
  \citenamefont {Weidem\"{u}ller},\ and\ \citenamefont
  {Grimm}}]{Schunemann1998}%
  \BibitemOpen
  \bibfield  {author} {\bibinfo {author} {\bibfnamefont {U.}~\bibnamefont
  {Sch\"{u}nemann}}, \bibinfo {author} {\bibfnamefont {H.}~\bibnamefont
  {Engler}}, \bibinfo {author} {\bibfnamefont {M.}~\bibnamefont
  {Zielonkowski}}, \bibinfo {author} {\bibfnamefont {M.}~\bibnamefont
  {Weidem\"{u}ller}}, \ and\ \bibinfo {author} {\bibfnamefont {R.}~\bibnamefont
  {Grimm}},\ }\href {http://dx.doi.org/10.1016/S0030-4018(98)00517-3}
  {\bibfield  {journal} {\bibinfo  {journal} {Opt. Commun.}\ }\textbf {\bibinfo
  {volume} {158}},\ \bibinfo {pages} {263} (\bibinfo {year}
  {1998})}\BibitemShut {NoStop}%
\bibitem [{\citenamefont {Lin}\ \emph {et~al.}(1991)\citenamefont {Lin},
  \citenamefont {Shimizu}, \citenamefont {Zhan}, \citenamefont {Shimizu},\ and\
  \citenamefont {Takuma}}]{Lin1991}%
  \BibitemOpen
  \bibfield  {author} {\bibinfo {author} {\bibfnamefont {Z.}~\bibnamefont
  {Lin}}, \bibinfo {author} {\bibfnamefont {K.}~\bibnamefont {Shimizu}},
  \bibinfo {author} {\bibfnamefont {M.}~\bibnamefont {Zhan}}, \bibinfo {author}
  {\bibfnamefont {F.}~\bibnamefont {Shimizu}}, \ and\ \bibinfo {author}
  {\bibfnamefont {H.}~\bibnamefont {Takuma}},\ }\href {\doibase
  10.1143/jjap.30.l1324} {\bibfield  {journal} {\bibinfo  {journal} {Japanese
  Journal of Applied Physics}\ }\textbf {\bibinfo {volume} {30}},\ \bibinfo
  {pages} {L1324} (\bibinfo {year} {1991})}\BibitemShut {NoStop}%
\bibitem [{\citenamefont {Fort}\ \emph {et~al.}(1998)\citenamefont {Fort},
  \citenamefont {Bambini}, \citenamefont {Cacciapuoti}, \citenamefont
  {Cataliotti}, \citenamefont {Prevedelli}, \citenamefont {Tino},\ and\
  \citenamefont {Inguscio}}]{Fort1998}%
  \BibitemOpen
  \bibfield  {author} {\bibinfo {author} {\bibfnamefont {C.}~\bibnamefont
  {Fort}}, \bibinfo {author} {\bibfnamefont {A.}~\bibnamefont {Bambini}},
  \bibinfo {author} {\bibfnamefont {L.}~\bibnamefont {Cacciapuoti}}, \bibinfo
  {author} {\bibfnamefont {F.~S.}\ \bibnamefont {Cataliotti}}, \bibinfo
  {author} {\bibfnamefont {M.}~\bibnamefont {Prevedelli}}, \bibinfo {author}
  {\bibfnamefont {G.~M.}\ \bibnamefont {Tino}}, \ and\ \bibinfo {author}
  {\bibfnamefont {M.}~\bibnamefont {Inguscio}},\ }\href {\doibase
  10.1007/s100530050154} {\bibfield  {journal} {\bibinfo  {journal} {Eur. Phys.
  J. D}\ }\textbf {\bibinfo {volume} {3}},\ \bibinfo {pages} {113} (\bibinfo
  {year} {1998})}\BibitemShut {NoStop}%
\bibitem [{\citenamefont {Xu}\ \emph {et~al.}(2003)\citenamefont {Xu},
  \citenamefont {Loftus}, \citenamefont {Dunn}, \citenamefont {Greene},
  \citenamefont {Hall}, \citenamefont {Gallagher},\ and\ \citenamefont
  {Ye}}]{Xu2003}%
  \BibitemOpen
  \bibfield  {author} {\bibinfo {author} {\bibfnamefont {X.}~\bibnamefont
  {Xu}}, \bibinfo {author} {\bibfnamefont {T.~H.}\ \bibnamefont {Loftus}},
  \bibinfo {author} {\bibfnamefont {J.~W.}\ \bibnamefont {Dunn}}, \bibinfo
  {author} {\bibfnamefont {C.~H.}\ \bibnamefont {Greene}}, \bibinfo {author}
  {\bibfnamefont {J.~L.}\ \bibnamefont {Hall}}, \bibinfo {author}
  {\bibfnamefont {A.}~\bibnamefont {Gallagher}}, \ and\ \bibinfo {author}
  {\bibfnamefont {J.}~\bibnamefont {Ye}},\ }\href {\doibase
  10.1103/physrevlett.90.193002} {\bibfield  {journal} {\bibinfo  {journal}
  {Physical Review Letters}\ }\textbf {\bibinfo {volume} {90}},\ \bibinfo
  {pages} {193002+} (\bibinfo {year} {2003})}\BibitemShut {NoStop}%
\bibitem [{\citenamefont {Dalibard}\ and\ \citenamefont
  {Cohen-Tannoudji}(1989)}]{Dalibard1989}%
  \BibitemOpen
  \bibfield  {author} {\bibinfo {author} {\bibfnamefont {J.}~\bibnamefont
  {Dalibard}}\ and\ \bibinfo {author} {\bibfnamefont {C.}~\bibnamefont
  {Cohen-Tannoudji}},\ }\href {\doibase 10.1364/JOSAB.6.002023} {\bibfield
  {journal} {\bibinfo  {journal} {J. Opt. Soc. Am. B}\ }\textbf {\bibinfo
  {volume} {6}},\ \bibinfo {pages} {2023} (\bibinfo {year} {1989})}\BibitemShut
  {NoStop}%
\bibitem [{\citenamefont {Guo}\ and\ \citenamefont {Berman}(1993)}]{Guo1993}%
  \BibitemOpen
  \bibfield  {author} {\bibinfo {author} {\bibfnamefont {J.}~\bibnamefont
  {Guo}}\ and\ \bibinfo {author} {\bibfnamefont {P.~R.}\ \bibnamefont
  {Berman}},\ }\href {http://dx.doi.org/10.1103/PhysRevA.48.3225} {\bibfield
  {journal} {\bibinfo  {journal} {Phys. Rev. A}\ }\textbf {\bibinfo {volume}
  {48}},\ \bibinfo {pages} {3225} (\bibinfo {year} {1993})}\BibitemShut
  {NoStop}%
\bibitem [{\citenamefont {Kerman}(2002)}]{Kerman2002}%
  \BibitemOpen
  \bibfield  {author} {\bibinfo {author} {\bibfnamefont {J.}~\bibnamefont
  {Kerman}},\ }\href@noop {} {Ph.D. thesis},\ \bibinfo  {school} {Stanford
  University} (\bibinfo {year} {2002})\BibitemShut {NoStop}%
\bibitem [{\citenamefont {Ketterle}\ \emph {et~al.}(1999)\citenamefont
  {Ketterle}, \citenamefont {Durfee},\ and\ \citenamefont
  {Stamper-Kurn}}]{Ketterle1999}%
  \BibitemOpen
  \bibfield  {author} {\bibinfo {author} {\bibfnamefont {W.}~\bibnamefont
  {Ketterle}}, \bibinfo {author} {\bibfnamefont {D.}~\bibnamefont {Durfee}}, \
  and\ \bibinfo {author} {\bibfnamefont {D.}~\bibnamefont {Stamper-Kurn}},\
  }in\ \href {http://arxiv.org/abs/cond-mat/9904034} {\emph {\bibinfo
  {booktitle} {Bose-Einstein Condensation in Atomic Gases}}}\ (\bibinfo
  {publisher} {IOS press},\ \bibinfo {address} {Amsterdam},\ \bibinfo {year}
  {1999})\ pp.\ \bibinfo {pages} {67--176}\BibitemShut {NoStop}%
\bibitem [{\citenamefont {Castin}\ \emph {et~al.}(1991)\citenamefont {Castin},
  \citenamefont {Dalibard},\ and\ \citenamefont
  {Cohen-Tannoudji}}]{Castin1990}%
  \BibitemOpen
  \bibfield  {author} {\bibinfo {author} {\bibfnamefont {Y.}~\bibnamefont
  {Castin}}, \bibinfo {author} {\bibfnamefont {J.}~\bibnamefont {Dalibard}}, \
  and\ \bibinfo {author} {\bibfnamefont {C.}~\bibnamefont {Cohen-Tannoudji}},\
  }in\ \href@noop {} {\emph {\bibinfo {booktitle} {Light Induced Kinetic
  Effects on Atoms, Ions, and Molecules}}},\ \bibinfo {editor} {edited by\
  \bibinfo {editor} {\bibfnamefont {L.}~\bibnamefont {Moi}}, \bibinfo {editor}
  {\bibfnamefont {S.}~\bibnamefont {Gozzini}}, \bibinfo {editor} {\bibfnamefont
  {C.}~\bibnamefont {Gabanini}}, \bibinfo {editor} {\bibfnamefont
  {E.}~\bibnamefont {Arimondo}}, \ and\ \bibinfo {editor} {\bibfnamefont
  {F.}~\bibnamefont {Strumia}}}\ (\bibinfo  {publisher} {ETS Editrice},\
  \bibinfo {address} {Pisa},\ \bibinfo {year} {1991})\ pp.\ \bibinfo {pages}
  {445--476}\BibitemShut {NoStop}%
\bibitem [{\citenamefont {Jersblad}\ \emph {et~al.}(2004)\citenamefont
  {Jersblad}, \citenamefont {Ellmann}, \citenamefont {St{\o}chkel},
  \citenamefont {Kastberg}, \citenamefont {Sanchez-Palencia},\ and\
  \citenamefont {Kaiser}}]{Jersblad2004}%
  \BibitemOpen
  \bibfield  {author} {\bibinfo {author} {\bibfnamefont {J.}~\bibnamefont
  {Jersblad}}, \bibinfo {author} {\bibfnamefont {H.}~\bibnamefont {Ellmann}},
  \bibinfo {author} {\bibfnamefont {K.}~\bibnamefont {St{\o}chkel}}, \bibinfo
  {author} {\bibfnamefont {A.}~\bibnamefont {Kastberg}}, \bibinfo {author}
  {\bibfnamefont {L.}~\bibnamefont {Sanchez-Palencia}}, \ and\ \bibinfo
  {author} {\bibfnamefont {R.}~\bibnamefont {Kaiser}},\ }\href {\doibase
  10.1103/physreva.69.013410} {\bibfield  {journal} {\bibinfo  {journal}
  {Physical Review A}\ }\textbf {\bibinfo {volume} {69}},\ \bibinfo {pages}
  {013410+} (\bibinfo {year} {2004})}\BibitemShut {NoStop}%
\bibitem [{\citenamefont {Kastberg}\ \emph {et~al.}(1995)\citenamefont
  {Kastberg}, \citenamefont {Phillips}, \citenamefont {Rolston}, \citenamefont
  {Spreeuw},\ and\ \citenamefont {Jessen}}]{Kastberg1995}%
  \BibitemOpen
  \bibfield  {author} {\bibinfo {author} {\bibfnamefont {A.}~\bibnamefont
  {Kastberg}}, \bibinfo {author} {\bibfnamefont {W.~D.}\ \bibnamefont
  {Phillips}}, \bibinfo {author} {\bibfnamefont {S.~L.}\ \bibnamefont
  {Rolston}}, \bibinfo {author} {\bibfnamefont {R.~J.~C.}\ \bibnamefont
  {Spreeuw}}, \ and\ \bibinfo {author} {\bibfnamefont {P.~S.}\ \bibnamefont
  {Jessen}},\ }\href {\doibase 10.1103/physrevlett.74.1542} {\bibfield
  {journal} {\bibinfo  {journal} {Physical Review Letters}\ }\textbf {\bibinfo
  {volume} {74}},\ \bibinfo {pages} {1542} (\bibinfo {year}
  {1995})}\BibitemShut {NoStop}%
\bibitem [{\citenamefont {Winoto}\ \emph {et~al.}(1999)\citenamefont {Winoto},
  \citenamefont {DePue}, \citenamefont {Bramall},\ and\ \citenamefont
  {Weiss}}]{Winoto1999}%
  \BibitemOpen
  \bibfield  {author} {\bibinfo {author} {\bibfnamefont {S.~L.}\ \bibnamefont
  {Winoto}}, \bibinfo {author} {\bibfnamefont {M.~T.}\ \bibnamefont {DePue}},
  \bibinfo {author} {\bibfnamefont {N.~E.}\ \bibnamefont {Bramall}}, \ and\
  \bibinfo {author} {\bibfnamefont {D.~S.}\ \bibnamefont {Weiss}},\ }\href
  {\doibase 10.1103/physreva.59.r19} {\bibfield  {journal} {\bibinfo  {journal}
  {Physical Review A}\ }\textbf {\bibinfo {volume} {59}},\ \bibinfo {pages}
  {R19} (\bibinfo {year} {1999})}\BibitemShut {NoStop}%
\bibitem [{\citenamefont {Grynberg}\ \emph {et~al.}(1993)\citenamefont
  {Grynberg}, \citenamefont {Lounis}, \citenamefont {Verkerk}, \citenamefont
  {Courtois},\ and\ \citenamefont {Salomon}}]{Grynberg1993}%
  \BibitemOpen
  \bibfield  {author} {\bibinfo {author} {\bibfnamefont {G.}~\bibnamefont
  {Grynberg}}, \bibinfo {author} {\bibfnamefont {B.}~\bibnamefont {Lounis}},
  \bibinfo {author} {\bibfnamefont {P.}~\bibnamefont {Verkerk}}, \bibinfo
  {author} {\bibfnamefont {J.~Y.}\ \bibnamefont {Courtois}}, \ and\ \bibinfo
  {author} {\bibfnamefont {C.}~\bibnamefont {Salomon}},\ }\href
  {http://dx.doi.org/10.1103/PhysRevLett.70.2249} {\bibfield  {journal}
  {\bibinfo  {journal} {Phys. Rev. Lett.}\ }\textbf {\bibinfo {volume} {70}},\
  \bibinfo {pages} {2249} (\bibinfo {year} {1993})}\BibitemShut {NoStop}%
\bibitem [{\citenamefont {Kerman}\ \emph {et~al.}(2000)\citenamefont {Kerman},
  \citenamefont {Vuleti\'{c}}, \citenamefont {Chin},\ and\ \citenamefont
  {Chu}}]{Kerman2000}%
  \BibitemOpen
  \bibfield  {author} {\bibinfo {author} {\bibfnamefont {A.~J.}\ \bibnamefont
  {Kerman}}, \bibinfo {author} {\bibfnamefont {V.}~\bibnamefont {Vuleti\'{c}}},
  \bibinfo {author} {\bibfnamefont {C.}~\bibnamefont {Chin}}, \ and\ \bibinfo
  {author} {\bibfnamefont {S.}~\bibnamefont {Chu}},\ }\href
  {http://dx.doi.org/10.1103/PhysRevLett.84.439} {\bibfield  {journal}
  {\bibinfo  {journal} {Phys. Rev. Lett.}\ }\textbf {\bibinfo {volume} {84}},\
  \bibinfo {pages} {439} (\bibinfo {year} {2000})}\BibitemShut {NoStop}%
\end{thebibliography}%

\end{document}